\documentclass[oneside,11pt]{article}

\usepackage[utf8]{inputenc}
\usepackage[T1]{fontenc}

\usepackage{amsthm}
\usepackage{amsmath}
\usepackage{amsfonts}
\usepackage{amssymb}

\usepackage{mathtools}

\usepackage{graphicx} 
\usepackage{float}
\usepackage{booktabs}
\usepackage{multirow}
\usepackage{rotating}
\usepackage{array}

\usepackage{breakcites}

\usepackage{enumitem}

\usepackage[top=1.5cm,bottom=2cm,right=2.5cm,left=2.5cm]{geometry}
\usepackage{titling}

\usepackage{natbib}

\usepackage{ifplatform}

\ifwindows
\fi

\ifwindows
	\usepackage{bbm}
\else
	\iflinux
		\usepackage{bbm}
	\else
		\usepackage{bbm}
	\fi
\fi

\newcommand{\lp}{\left(}
\newcommand{\rp}{\right)}

\newcommand{\lb}{\left\{}
\newcommand{\rb}{\right\}}

\newcommand{\R}{\mathbb{R}}
\newcommand{\Z}{\mathbb{Z}}
\newcommand{\lrc}[1]{\left[#1\right]}

\DeclareFontFamily{OT1}{pzc}{}
\DeclareFontShape{OT1}{pzc}{m}{it}{<-> s * [1.10] pzcmi7t}{}
\DeclareMathAlphabet{\mathpzc}{OT1}{pzc}{m}{it}

\usepackage[pdftex,final,bookmarksnumbered,bookmarksopen=false,breaklinks,colorlinks]{hyperref}
\hypersetup{
	final,
    bookmarksnumbered=true,
    bookmarksopen=true,
    bookmarksopenlevel=0,
    unicode=false,          
    pdftoolbar=true,        
    pdfmenubar=true,        
    pdffitwindow=false,     
    pdftitle={Exploring wind direction and SO2 concentration by circular-linear density estimation},    
	pdfdisplaydoctitle=true, 
	pdftoolbar=true,
	pdfmenubar=true,
	pdflang={English},
    pdfauthor={Eduardo Garcia-Portugues, Rosa M. Crujeiras, Wenceslao Gonzalez-Manteiga},     
    pdfsubject={arXiv paper},   
    pdfcreator={Eduardo Garcia-Portugues},   
    pdfproducer={Eduardo Garcia-Portugues}, 
    pdfkeywords={Circular distributions} {Circular kernel estimation} {Circular-linear data} {Copula}, 
    pdfnewwindow=true,      
    breaklinks=true,
    hidelinks,       
    linkcolor=black,        
    citecolor=black,        
    filecolor=magenta,      
    urlcolor=cyan           
}

\newtheorem{algo}{Algorithm}
\newtheorem{example}{Example}

\allowdisplaybreaks

\begin{document}

\title{Exploring wind direction and SO$_2$ concentration by circular-linear density estimation}
\setlength{\droptitle}{-1cm}
\predate{}%
\postdate{}%
\author{Eduardo Garc\'ia-Portugu\'es$^{1,2}$, Rosa M. Crujeiras$^1$, and Wenceslao Gonz\'alez-Manteiga$^1$}

\date{}

\footnotetext[1]{
Department of Statistics and Operations Research, University of Santiago de Compostela (Spain).}
\footnotetext[2]{Corresponding author. e-mail: \href{mailto:eduardo.garcia@usc.es}{eduardo.garcia@usc.es}.}

\maketitle


\begin{abstract}
The study of environmental problems usually requires the description of variables with different nature and the assessment of relations between them. In this work, an algorithm for flexible estimation of the joint density for a circular-linear variable is proposed. The method is applied for exploring the relation between wind direction and SO$_2$ concentration in a monitoring station close to a power plant located in Galicia (NW-Spain), in order to compare the effectiveness of precautionary measures for pollutants reduction in two different years.
\end{abstract}

\begin{flushleft}
\end{flushleft}

\section{Introduction}
\label{copcirlin:introduction}

Air pollution studies require the investigation of relationships between emission sources and pollutants concentration in nearby sites. In addition, the effectiveness of environmental policies with the aim of pollution reduction should be checked, at least in a descriptive way, in order to assess whether the implemented precautionary measurements had worked out.\\

Different statistical methods have been considered for the study of the relation between wind direction and pollutants concentration, both for exploratory and for inferential analysis, taking into account that wind direction is a circular variable requiring a proper statistical treatment. Wind potential assessment using descriptive methods and spectral analysis has been carried out by \cite{Shih2008}. In addition, wind direction has been proved to play a significant role in detection of emission sources (see \cite{Chen2011}) and air quality studies (see \cite{Bayraktar2010}), although the wind direction is not treated as a circular variable, but discretized as a factor. \cite{Jammalamadaka2006} considered regression models for the pollutants concentration (linear response) over the wind direction (circular explanatory variable), constructing the regression function in terms of the sine and cosine components of the circular variable. Recently, \cite{Deschepper2008} introduced a graphical diagnostic tool, jointly with a test, for fit assessment in parametric circular-linear regression, illustrating the technique in an air quality environmental study.\\

From a more technical perspective, \cite{Johnson1978} and \cite{Wehrly1980} presented a method for obtaining joint circular-linear and circular-circular densities with specified marginals, respectively. \cite{Fern'andez-Dur'an2004} introduced a new family of circular distributions based on nonnegative trigonometric sums, and this idea is used in \cite{Fern'andez-Dur'an2007} in the construction of circular-linear densities, adapting the proposal of \cite{Johnson1978}. The introduction of nonnegative trigonometric sums for modelling the circular distributions involved in the formulation of \cite{Johnson1978} allows for more flexible models, that may present skewness or multimodality, features that cannot be reflected through the von Mises distribution (the classical model for circular variables). However, the flexibility\nopagebreak[4] claimed in this\nopagebreak[4] proposal can be also obtained by a completely nonparametric approach. \\

In this work, a procedure for modelling the relation between a circular and a linear variable is proposed. The relation is specified by a circular-linear density which, represented in terms of copulas, can be estimated nonparametrically. The estimation algorithm can be also adapted to a semiparametric framework, when an underlying model for the marginal distributions can be imposed, and it also comprises the classical \cite{Johnson1978} model. It also enables the construction of an estimation framework without imposing an underlying parametric model. The copula approach presents some computational advantages, in order to carry out a simulation study.\\

\begin{figure}[h!]
\centering
\includegraphics[scale=0.45]{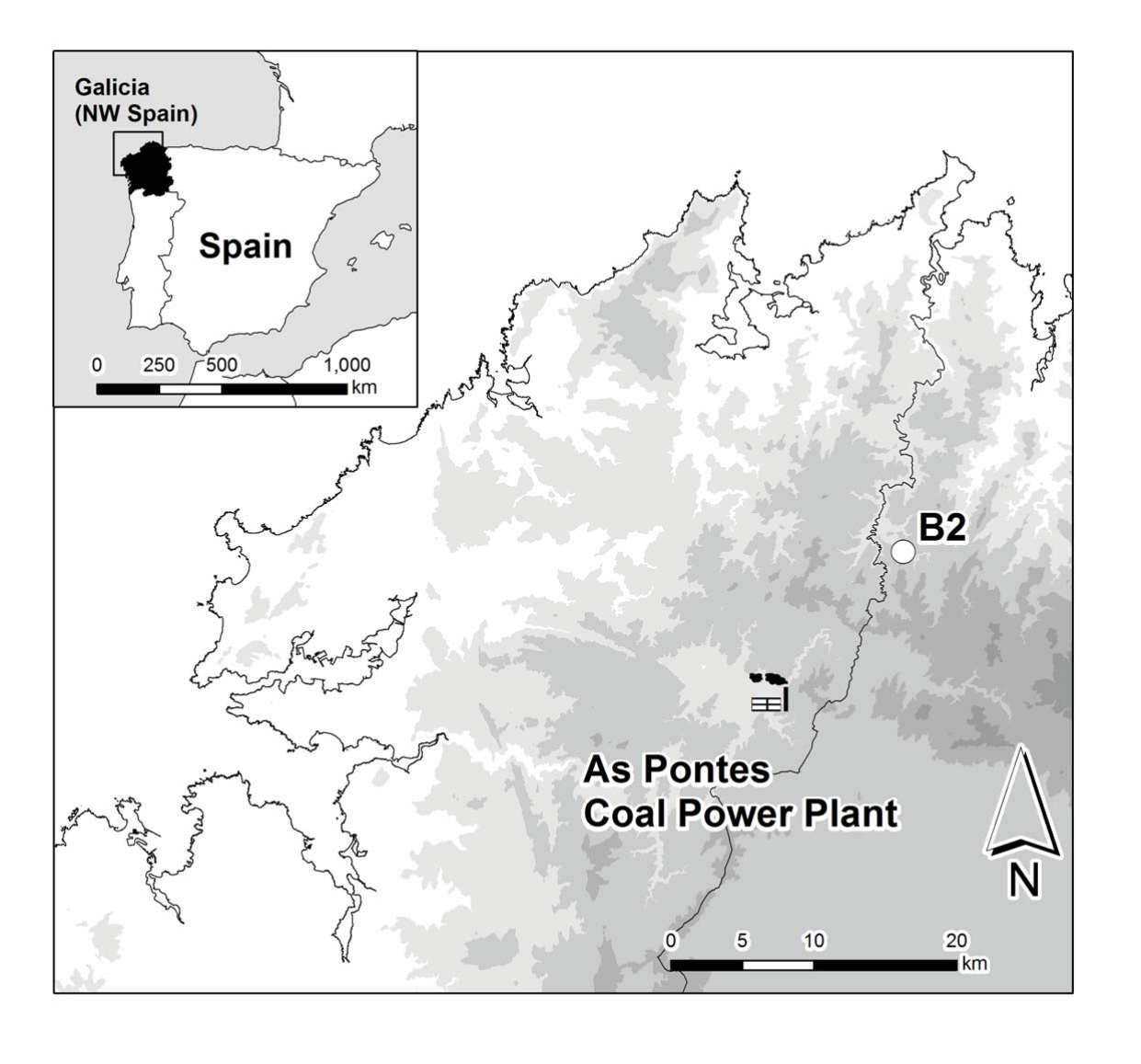}
\caption{\small Locations of monitoring station (circle) and power plant in Galicia (NW-Spain). Location of station B2: $7^{\circ}$ $44$' $10$'' W, $43^\circ$ $32$' $05$'' N. Power plant location: $7^\circ$ $51$' $45$'' W, $43^\circ$ $26$' $26$'' N.}
\label{copcirlin:locations}
\end{figure}

The practical aim of this work is to explore the relation between wind incidence direction (wind blowing from this direction) and SO$_2$ levels in a monitoring station close to a power plant located in Galicia (NW-Spain). The monitoring station and the thermal power plant locations  are shown in Figure \ref{copcirlin:locations}. In the power plant, energy was usually produced from the combustion of local coal, which also generates pollutants as sulphur dioxide (SO$_2$). In order to reduce the emission of SO$_2$ to the atmosphere, and to comply with European regulations, coal with less sulphur content has been used since 2005. In addition, the power plant changed the energy production system by settling a combined process of coal and gas burning in 2008. These measures were aimed to reduce the SO$_2$ emissions.

\begin{figure}[h!]
\centering
\includegraphics[width=14cm]{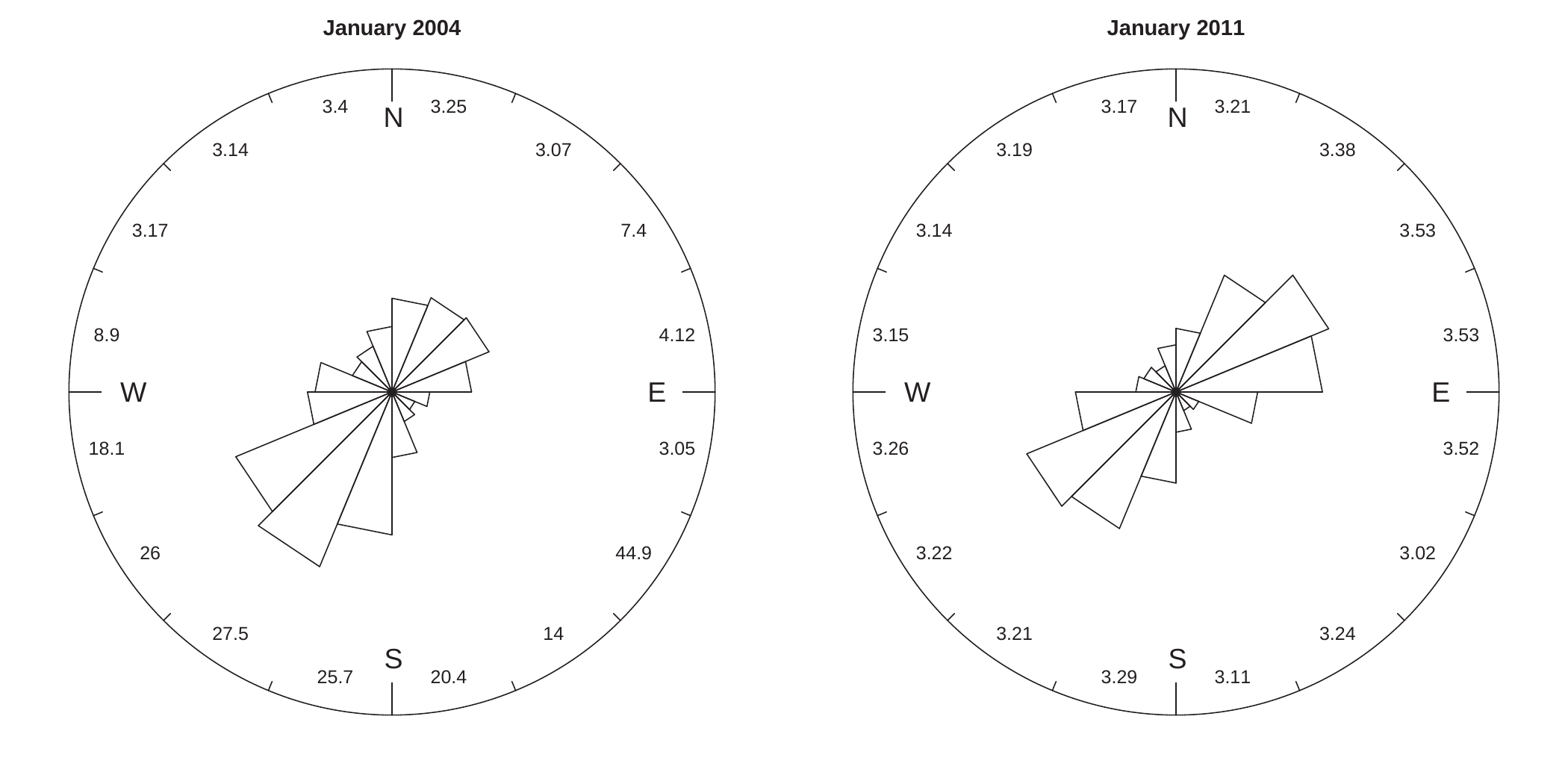}
\caption{\small Rose diagrams for wind direction in station B2 for 2004 (left plot) and 2011 (right plot), with average SO$_2$ concentration.}
\label{copcirlin:diagrams}
\end{figure}

The analysed data corresponds to SO$_2$ and wind incidence direction measurements taken during January 2004 and January 2011, with one minute frequency. The monitoring station B2 (see Figure \ref{copcirlin:locations}) is located in a wind farm $13.4$ kilometres in the NE direction with respect to the power plant. In Figure \ref{copcirlin:diagrams}, rose diagrams for wind direction are shown, including also the average SO$_2$ concentration for each wind direction sector. It can be seen that higher values of SO$_2$ are shown in 2004. Note also that there are two dominant wind incidence directions, specifically, blowing from SW and from NE. In addition, note that the average SO$_2$ values for winds blowing from SW are remarkably high for 2004, which is explained by the position of the monitoring station with respect to the power plant. However, the relation between wind direction and SO$_2$ levels is not clear from these representations, and the dependency (or lack of dependency) between them should be investigated. \\

This work is organized as follows. Section \ref{copcirlin:background} provides some background on circular and circular-linear random variables and a brief review on copula methods. The algorithm for estimating a circular-linear density is detailed and discussed in Section \ref{copcirlin:estimation}. The finite sample properties of the algorithm, combining parametric and nonparametric approaches, are illustrated by a simulation study. The completely nonparametric version of this algorithm is applied for analysing the relation between wind direction and SO$_2$ concentration in Section \ref{copcirlin:application}. Some final comments are given in Section \ref{copcirlin:final}.

\section{Background}
\label{copcirlin:background}

The main goal of this work is to describe the relation between wind direction and SO$_2$ concentration in a monitoring station close to a power plant at two different time moments, before and after precautionary measurements to reduce SO$_2$ emissions were applied. Bearing in mind the different nature of the variables and noticing that measurements from wind direction are angles, some background on circular and circular-linear random variables is introduced. This methodology will be needed in order to describe the wind direction itself and the joint relation between the two variables. For that purpose, the \cite{Johnson1978} family (J\&W in what follows) of circular-linear distributions will be introduced. A general procedure, based on the copula representation of a density, allows for a more flexible estimation framework. The goal of this copula representation is twofold: firstly, the classical J\&W family can be written in such a way, just involving univariate (circular and linear) densities; secondly, with the copula representation, flexible circular-linear relations beyond this specific model are also possible.

\subsection{Some circular and circular-linear distributions}

Denote by $\Theta$ a circular random variable with support in the unit circle $\mathbb S^1$. A circular distribution $\Psi$ for $\Theta$ assigns a probability to each direction $(\cos(\theta),\sin(\theta))$ in the plane $\R^2$, characterized by the angle $\theta\in[0,2\pi)$ (see \cite{Mardia2000} for a survey on statistical methods for circular data). The von Mises distribution is the analogue of the normal distribution in circular random variables. This family of distributions, usually denoted by $\mathrm{vM}(\mu,\kappa)$, is characterized by two parameters: $\mu\in[0,2\pi)$, the circular mean and $\kappa\geq0$, a circular concentration parameter around $\mu$. The corresponding density function is given by
\begin{equation}
\varphi_{\mathrm{vM}}(\theta;\mu,\kappa)=\frac{1}{2\pi \mathcal{I}_0(\kappa)}e^{\kappa\cos(\theta-\mu)},\quad \theta\in[0,2\pi),
\label{copcirlin:vonMises}
\end{equation}
being $\mathcal{I}_0(\kappa)=\frac{1}{2\pi}\int_0^{2\pi}e^{\kappa\cos\omega}d\omega$ the modified Bessel function of first kind and order zero. The von Mises cumulative distribution function, considering the zero angle as the starting point, is defined \nolinebreak[4]as:
\[
\Psi_{\mathrm{vM}}(\theta;\mu,\kappa)=\int_0^{\theta}\varphi_{\mathrm{vM}}(\omega;\mu,\kappa)\,d\omega,\quad\theta\in[0,2\pi).
\]
The uniform circular distribution, 
\begin{equation}
\varphi_\mathrm{U}(\theta)=\frac{1}{2\pi},\quad \theta\in[0,2\pi),
\label{copcirlin:uniform}
\end{equation}
is obtained as a particular case of the von Mises family, for $\kappa=0$. Circular density estimation can be performed by parametric methods, such as Maximum Likelihood Estimation (MLE), or using nonparametric techniques, based on kernel approaches, as will be seen later.\\

In order to explain the relation between a circular and a linear random variable (in our example, wind direction and SO$_2$ concentration), the construction of a joint circular-linear density will be considered. A circular-linear random variable $(\Theta,X)$ is supported on the cylinder $\mathbb{S}^1\times \R$ or in a subset of it and a circular-linear density for $(\Theta,X)$, namely $p$, must satisfy a periodicity condition in the circular argument, that is:
\[
p(\theta,x)=p(\theta+2\pi k,x),\quad \theta\in[0,2\pi),\,x\in\R,\,k\in\Z,
\]
as well as the usual assumptions on taking nonnegative values and integrating one. \cite{Johnson1978} proposed a method for obtaining circular-linear densities with specified marginals. Denote by $\varphi$ and $f$ the circular and linear marginal densities, respectively, and by $\Psi$ and $F$ their corresponding distribution functions. Let also $g$ be another circular density. Then
\begin{equation}
p(\theta,x)=2\pi g\lp 2\pi\lp \Psi(\theta)-F(x) \rp\rp\times \varphi(\theta)f(x)	
\label{copcirlin:circular_linear_density}
\end{equation}
is a density for a circular-linear distribution for a random variable $(\Theta,X)$, with specified marginal densities $\varphi$ and $f$  (see \cite{Johnson1978}, Theorem 5). Circular-linear densities with specified marginals can be also obtained considering the sum of the marginal distributions in the argument of the joining density in (\ref{copcirlin:circular_linear_density}). Circular-linear densities may include von Mises and Gaussian marginals, but the dependence between them will be specified by the joining density $g$. In fact, the independence model corresponds to taking $\varphi_U$ in (\ref{copcirlin:uniform}) as the joining density. From a data sample of $(\Theta,X)$, assuming that the joint density can be represented as in (\ref{copcirlin:circular_linear_density}), an estimator of $p$ could be obtained by the estimations of the marginals and the joining density. \cite{Wehrly1980} proved that the construction of circular-circular distributions (that is, distributions on the torus) can be done similarly to (\ref{copcirlin:circular_linear_density}), just considering prespecified circular marginal distributions. Note that (\ref{copcirlin:circular_linear_density}) is just a construction method and not a characterization of circular-linear densities. In addition, there are no available testing procedures for checking if a certain dataset follows such a distribution. Hence, a more general approach for circular-linear density construction would be helpful.\\

In the next section, some background on copulas will be introduced, allowing for a more flexible procedure for obtaining circular-linear densities with specified marginals, where the representation in (\ref{copcirlin:circular_linear_density}) fits as a particular case. With this proposal, a fully nonparametric estimation procedure can be applied. 

\subsection{Some notes on copulas}
\label{copcirlin:copulas}

Copula functions are multivariate distributions with uniform marginals (see \cite{Nelsen2006} for a complete review on copulas). One of the main results in copula theory is Sklar's theorem, which, in the bivariate case, states that if $F$ is a joint distribution function with marginals $F_1$ and $F_2$ then there exists a copula $C$ such that:
\begin{equation}
F(x,y)=C(F_1(x),F_2(y)), \quad x,y\in \R. 
\label{copcirlin:copula}
\end{equation}
If $F_1$ and $F_2$ are continuous distributions, then $C$ is unique. Conversely, if $C$ is a copula and $F_1$ and $F_2$ are distribution functions, then $F$ defined by (\ref{copcirlin:copula}) is a bivariate distribution with\nopagebreak[4] marginals $F_1$ and $F_2$.\\

If the marginal random variables are absolutely continuous, Sklar's result can be interpreted in terms of the corresponding densities. Denoting by $c$ the copula density, the bivariate density of $F$ in (\ref{copcirlin:copula}) can be written as
\[
f(x,y)=c(F_1(x),F_2(y))\times f_1(x)f_2(y),\quad x,y\in \R. 
\]

As pointed out by \cite{Nazemi2012}, copula modelling provides a simple but powerful way for describing the interdependence between environmental variables. In our particular setting, the nature of the variables is different, being the variables of interest linear (SO$_2$ concentration) and circular (wind direction). Circular-linear copulas, that will be denoted by $C_{\Theta,X}$, can be also defined taking into account the characteristics of the circular marginal, satisfying $c_{\Theta,X}(0,v)=c_{\Theta,X}(1,v)$, $\forall v\in[0,1]$, where $c_{\Theta,X}$ is the corresponding circular-linear copula density. Hence, a circular-linear density with marginals $\varphi$ and $f$ is given by:
\begin{equation}
p(\theta,x)=c_{\Theta,X}(\Psi(\theta),F(x))\times\varphi(\theta)f(x),\quad\theta\in[0,2\pi),\;x\in\mathbb R.
\label{copcirlin:circular_linear_copula}
\end{equation}

Note that J\&W's proposal can be seen as a particular case of (\ref{copcirlin:circular_linear_copula}). For a certain joining density $g$ in (\ref{copcirlin:circular_linear_density}), the corresponding copula density is given by:
\begin{equation}
c_{\Theta,X}\lp\Psi(\theta),F(x)\rp=2\pi g\lp2\pi\lp\Psi(\theta)\pm F(x)\rp\rp,
\label{copcirlin:cg_link}
\end{equation}
where the sign $\pm$ refers to the possibility of considering the sum or the difference of the marginal distributions in the argument of $g$, as it was previously mentioned.\\

\begin{figure}[h!]
\centering
\includegraphics[width=15cm]{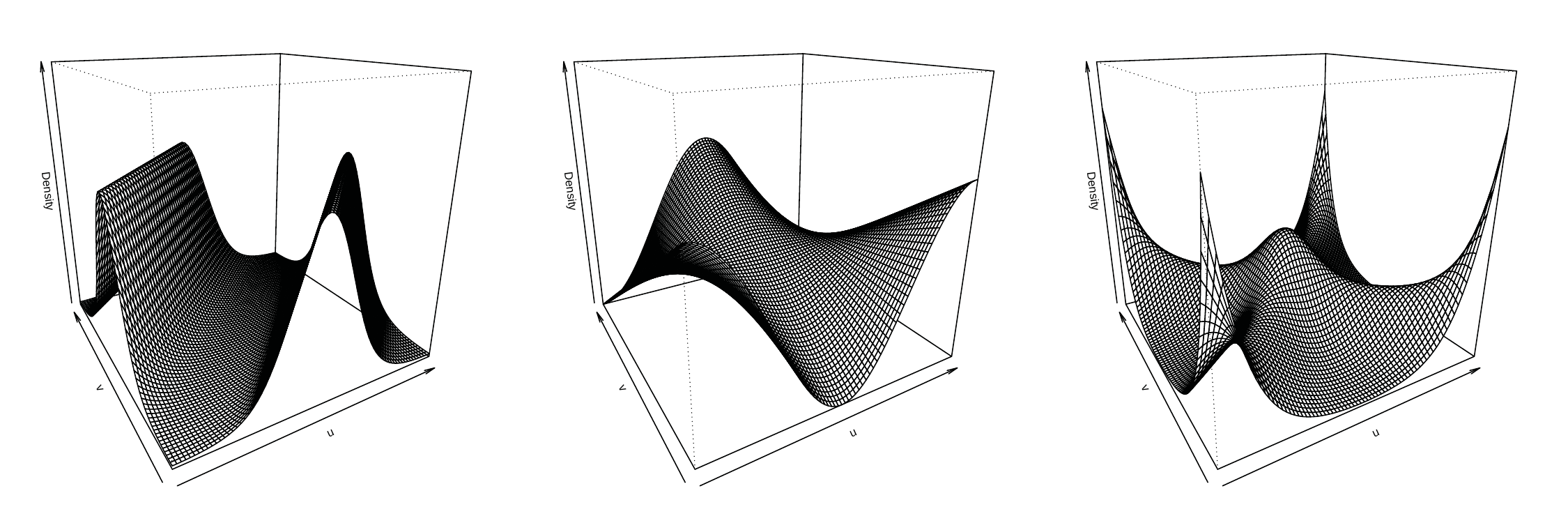}
\caption{\small Copula density surfaces. Left plot: J\&W copula with von Mises joining density with parameters $\mu=\pi$ and $\kappa=2$. Middle plot: QS-copula with $\alpha=(2\pi)^{-1}$. Right plot: reflected Frank copula with $\alpha=10$.}
\label{copcirlin:copula_examples}
\end{figure}

A circular-circular density, following the result of \cite{Wehrly1980}, can be constructed similarly just considering circular marginals for $(\Theta,\Omega)$ and guaranteeing that the copula density satisfies $c_{\Theta,\Omega}(0,v)=c_{\Theta,\Omega}(1,v)$ and $c_{\Theta,\Omega}(u,0)=c_{\Theta,\Omega}(u,1)$, $\forall u,v\in[0,1]$. For simplicity, the copula density in (\ref{copcirlin:cg_link}) will be denoted as J\&W copula (see Figure \ref{copcirlin:copula_examples}, left plot). \\

The representation of a circular-linear density in (\ref{copcirlin:circular_linear_copula}) enables the construction of new families of circular-linear distributions. From Corollary 3.2.5. in \cite{Nelsen2006}, a new family of circular-linear copulas with quadratic section (QS copula) in the linear component can be constructed. The copula densities in this family are given by
\begin{equation}
c_{\Theta,X}^{\alpha}(u,v)=1+2\pi\alpha\cos(2\pi u)(1-2v),\quad|\alpha|\leq(2\pi)^{-1}.
\label{copcirlin:copula_new_family}
\end{equation}

The QS copula family is parametrized by $\alpha$, which accounts for the deviation from the independence copula corresponding to $\alpha=0$. Figure \ref{copcirlin:copula_examples} (middle plot) shows the wavy surface corresponding to $\alpha=(2\pi)^{-1}$. The position of the three modes in the density, centred along $u=0$, $u=0.5$ and $u=1$, as well as their concentration, is controlled by the value of $\alpha$.  \\

A possible way to derive new copulas is through mixtures of other copulas (see \cite{Nelsen2006}). Thus, for any copula $\tilde c$, the mixture 
\begin{equation}
c_{\Theta,X}(u,v)=\frac{1}{4}\lp \tilde c(u,v)+\tilde c(1-u,v)+ \tilde c(u,1-v)+\tilde c(1-u,1-v) \rp,
\label{copcirlin:copula_new_family2}
\end{equation}
leads to a new copula satisfying  $c_{\Theta,X}(0,v)=c_{\Theta,X}(1,v)$, $\forall v\in[0,1]$. This construction also satisfies $c_{\Theta,X}(u,0)=c_{\Theta,X}(u,1)$, $\forall u\in[0,1]$, and provides a circular-circular copula (which is also circular-linear). A parametrized copula density $c_{\Theta,X}^{\alpha}$ can be obtained considering, for example, the parametric Frank copula:
\[
\tilde c_{\alpha}(u,v)=\frac{\alpha(1-e^{-\alpha})e^{-\alpha(u+v)}}{\lp(1-e^{-\alpha})-(1-e^{-\alpha u})(1-e^{-\alpha v})\rp^2},\quad \alpha\neq0.
\]
The mixture copula (\ref{copcirlin:copula_new_family2}) will be referred to as the \textit{reflected} copula of $\tilde c$. The parameter $\alpha$ also measures the deviation from independence, which is a limit case as $\alpha$ tends to zero. The copula density surface can be seen in Figure \ref{copcirlin:copula_examples} (right plot). In this example, the copula density surface shows five modes concentrated in the corners and the middle point of the unit square, and the peakedness of the modes increases as $\alpha$ grows.\\

These three families will be considered in the simulation study. It should be also mentioned that the copula representation poses some computational advantages in order to reproduce by simulation data samples from circular-linear distributions (see Section \ref{copcirlin:simul}). Finally, although this work is focused on the circular-linear case, some comments will be also made about circular-circular distributions.

\section{Estimation algorithm}
\label{copcirlin:estimation}

Denote by $\{\left(\Theta_i,X_i\right)\}_{i=1}^n$ a random sample of data for the circular-linear random variables $(\Theta,X)$ and consider the copula representation for $p$ in (\ref{copcirlin:circular_linear_copula}). In this joint circular-linear density model, three density functions must be estimated: the marginal densities $\varphi$ and $f$ (and also the corresponding distributions) and the copula density $c_{\Theta,X}$. A new natural procedure for estimating $p$ is given in the following algorithm.

\begin{algo}[Estimation algorithm]
\label{copcirlin:algo:1}
\mbox{}
\begin{enumerate}[label=\textit{\roman{*}}., ref=\textit{\roman{*}}]
	\item Obtain estimators for the marginal densities $\hat \varphi$, $\hat f$ and the corresponding marginal distributions $\hat \Psi$, $\hat F$.\label{copcirlin:algo:1:1}
	\item Compute an artificial sample $\big\{\big(\hat\Psi\big(\Theta_i\big),\hat F\big(X_i\big)\big)\big\}_{i=1}^n$ and estimate the copula density $\hat c_{\Theta,X}$.\label{copcirlin:algo:1:2}
	\item Obtain the circular-linear density estimator as $\hat p\lp\theta,x\rp=\hat c_{\Theta,X}(\hat\Psi(\theta),\hat F(x))\times\hat \varphi(\theta)\hat f(x)$.\label{copcirlin:algo:1:3}
\end{enumerate}
\end{algo}

The estimation of the marginal densities in step \ref{copcirlin:algo:1:1} can be done by parametric methods or by nonparametric procedures. For instance, a parametric estimator for $\hat f$ (respectively, for $\hat F$) can be obtained by MLE. In the circular case, that is, for obtaining $\hat\varphi$, MLE approaches are also possible (see \cite{Jammalamadaka2001}, Chapter 4).  These estimators are consistent, although restricted to parametric families such as the von Mises distribution or mixtures of von Mises. \\

Nonparametric kernel density estimation for linear random variables was introduced by Parzen and Rosenblatt (see \cite{Wand1995} for references on kernel density estimation) and the properties of this estimator have been well studied in the statistical literature. Consider $\{X_i\}_{i=1}^n$ a random sample of a linear variable $X$ with density $f$. The kernel density estimator of $f$ in a point $x\in\R$ is given by
\begin{equation}
\hat f_h(x)=\frac{1}{nh}\sum_{i=1}^nK\left(\frac{x-X_i}{h}\right),
\label{copcirlin:kernel_linear}
\end{equation}
where $K$ is a kernel function (usually a symmetric and unimodal density) and $h$ is the bandwidth parameter. One of the crucial problems in kernel density estimation is the bandwidth choice. There exist several alternatives for obtaining a global bandwidth minimizing a certain error criterion, usually the Mean Integrated Squared Error (MISE), such as the rule-of-thumb, least-squares cross-validatory procedures (see \cite{Wand1995}) or other plug-in rules, like the one proposed by \cite{Sheather1991}.\\

\cite{Hall1987} introduced a nonparametric kernel density estimator for directional data in the $q$-dimensional sphere $\mathbb{S}^q$. For the circular case ($q=1$), denoting by $\Theta$ a random variable with density $\varphi$, the circular kernel density estimation from a sample $\lb\Theta_i\rb_{i=1}^n$ is given by
\begin{equation}
\hat{\varphi}_\nu(\theta)=\frac{c_0(\nu)}{n}\sum_{i=1}^n L\lp\nu\cos(\theta-\Theta_i)\rp,\quad \theta\in[0,2\pi),
\label{copcirlin:kernel_circular}
\end{equation}
where $L$ is the circular kernel, $\nu$ is the circular bandwidth and $c_0(\nu)$ is a constant such that $\hat{\varphi}_\nu$ is a density. Some differences should be noted in contrast to the linear kernel density estimator in (\ref{copcirlin:kernel_linear}). First, the kernel function $L$ must be a rapidly varying function, such as the exponential (see \cite{Hall1987}). Secondly, the behaviour of $\nu$ is opposite to $h$: in linear kernel density estimation, small values of the bandwidth $h$ produce undersmoothed estimators (small values of $\nu$ oversmooth the density), whereas large values of $h$ give oversmoothed curves (large values of $\nu$ produce undersmoothing). See \cite{Hall1987} for a detailed description of the estimator and its properties.\\

As in the linear case, bandwidth selection is also an issue in circular kernel density estimation. Although in the linear case it is a well-studied problem, for circular density estimation there are still some open questions. \cite{Hall1987} proposed selecting the smoothing parameter by maximum likelihood cross-validation. There are other recent proposals, such as the automatic bandwidth selection method introduced by \cite{Taylor2008}, but based on his results none of the\nopagebreak[4] selectors proposed seems to show a superior behaviour. \\

Although for the marginal distributions it may be reasonable to assume a parametric model, it is not that clear for the copula function, regarding the dependence structure between the marginals. Hence, in a general situation, the copula estimation in step \ref{copcirlin:algo:1:2} should be carried out by a nonparametric procedure that will be explained below. However, for the J\&W density in (\ref{copcirlin:circular_linear_density}), the copula density $c_{\Theta,X}$ is linked with a joining circular density $g$ in (\ref{copcirlin:cg_link}) and this circular density can be estimated in the same way as the marginal circular density. Note that, in this family, all the estimators involved in the algorithm are obtained in a strictly univariate way, which simplifies their computation. \\

\begin{figure}[h!]
\centering
\includegraphics[width=14cm]{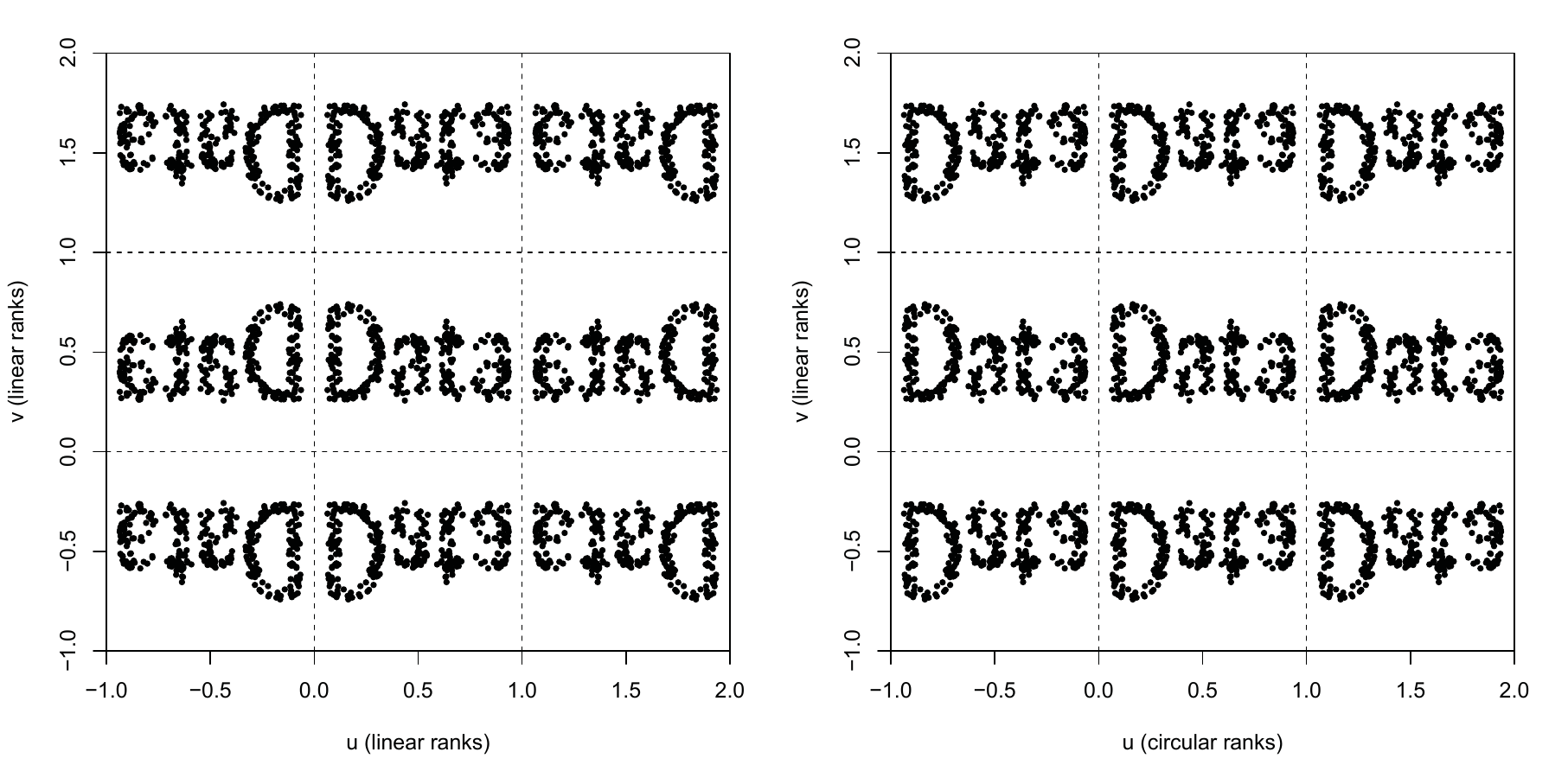}
\caption{\small Illustration for data reflection for copula density estimation. The central square in each plot corresponds with the original data ranks. Left plot: mirror reflection from \cite{Gijbels1990}. Right plot: circular-mirror reflection for estimator (\ref{copcirlin:np_copula}). }
\label{copcirlin:reflection}
\end{figure}

Nonparametric copula density estimation can be also done by kernel methods, as proposed by \cite{Gijbels1990}. The proposed estimator is similar to the classical bivariate kernel density estimator, with a product kernel and with a mirror image data modification. This mirror image (see Figure \ref{copcirlin:reflection}, left plot) consists in reflecting the data with respect to all edges and corners of the unit square, in order to reduce the edge effect. In our particular case of circular-linear copula densities, the reflection must be done accounting for the circular nature of the first component, as shown in Figure \ref{copcirlin:reflection}, right plot.

The copula density kernel estimator can be defined as:
\begin{equation}
\hat c_{\Theta, X}(u,v)=\frac{1}{n}\sum_{i=1}^n\sum_{l=1}^9\tilde{K}_B\left(u-\hat\Psi\Big(\Theta_i^{(l)}\Big),v-\hat F\Big(X_i^{(l)}\Big)\right),
\label{copcirlin:np_copula}
\end{equation}
where $\hat\Psi$ and $\hat F$ are the marginal kernel distribution functions, $\tilde K_B(u,v)=|B|^{-1}\tilde K(B^{-1/2}(u,v))$ is a bivariate rescaled kernel (see \cite{Ruppert1994} for notation) and $B$ is the bandwidth matrix ($|B|$ denotes the determinant). A plug-in rule for selecting the bandwidth matrix $B$ was proposed by \cite{Duong2003}. The reflected data sample in each square $l=1,\ldots,9$, namely $\big\{\big(\hat \Psi\big(\Theta_i^{(l)}\big),\hat F\big(X_i^{(l)}\big)\big)\big\}_{i=1}^n$ is obtained by rows in the $3\times 3$ plot (Figure \ref{copcirlin:reflection}, right plot) as follows: $(u-1,-v), (u,-v), (u+1,-v)$ (bottom row); $(u-1,v), (u,v), (u+1,v)$ (middle row);\nopagebreak[4] $(u-1,2-v), (u, 2-v), (u+1,2-v)$ (top row).\\

As pointed out by \cite{Omelka2009}, the estimator proposed by \cite{Gijbels1990} in the linear case still suffers from corner bias, which could be corrected by bandwidth shrinking. This issue is not so important in the circular-linear setting, given the periodicity condition. It should be also noticed that, with this construction, the estimator obtained in step \ref{copcirlin:algo:1:3}, with the proposed reflection, is guaranteed to be a density as long as the distribution and density estimators satisfy that $\hat F'=\hat f$ and $\hat\Psi'=\hat\varphi$.\\

The estimation algorithm can be extended for circular-circular densities, just with suitable (circular) density estimators for the marginals and a slight modification of the data reflection for the copula estimation. Specifically, reflection in $3\times 3$ scheme as the one shown in Figure \ref{copcirlin:reflection}, the middle one corresponding to the original circular-circular data quantiles will be done as follows: $(u-1,v), (u,v), (u+1,v)$ (bottom row); $(u-1,v), (u,v), (u+1,v)$ (middle row); $(u-1,v), (u, v), (u+1,v)$ (top row).

\subsection{Some simulation results}
\label{copcirlin:simul}

In order to check the performance of the estimation algorithm for circular-linear densities, the following scenarios are reproduced. Examples \ref{copcirlin:p_example1} and \ref{copcirlin:p_example2} were proposed by \cite{Johnson1978}. Example \ref{copcirlin:p_example3} corresponds to the QS-copula family given by (\ref{copcirlin:copula_new_family}) and Example \ref{copcirlin:p_example4} to the reflected Frank copula constructed in (\ref{copcirlin:copula_new_family2}).

\begin{example}[J\&W copula with circular uniform and normal marginal distributions]
\label{copcirlin:p_example1}
Let $\varphi_\mathrm{U}$ denote the circular uniform density (\ref{copcirlin:uniform}) and $\phi$ the standard normal density ($\Phi$ the standard normal distribution). Take the joining density $g=\varphi_{\mathrm{vM}}(\cdot;\mu,\kappa)$. A circular-linear density with marginals $\varphi$ and $\phi$ is given by
\[
p_1(\theta,x)=\frac{1}{\mathcal{I}_0(\kappa)}\exp\lb \kappa\cos(\theta-2\pi\Phi(x)-\mu)\rb\times\varphi_\mathrm{U}(\theta)\phi(x).
\]
\end{example}

\begin{example}[J\&W copula with von Mises and normal marginal distributions]
\label{copcirlin:p_example2}
Consider $g=\varphi_{\mathrm{vM}}(\cdot;\mu',\kappa')$ and the density marginals $\phi$ and $\varphi_{\mathrm{vM}}(\cdot;\mu_2,\kappa_2)$ (with corresponding distributions $\Phi$ and $\Psi_{\mathrm{vM}}(\cdot;\mu_2,\kappa_2)$, respectively). A joint circular-linear density is given by
\[
p_2(\theta,x)=\frac{1}{\mathcal{I}_0(\kappa')}\exp\lb \kappa'\cos\lp2\pi(\Psi_{\mathrm{vM}}(\theta;\mu_2,\kappa_2)-\Phi(x))-\mu'\rp\rb\times\varphi_{\mathrm{vM}}(\theta;\mu_2,\kappa_2)\phi(x).
\]
\end{example}

\begin{example}[QS-copula with von Mises and normal marginal distributions]\label{copcirlin:p_example3}
Take $\varphi_{\mathrm{vM}}(\cdot;\mu_3,\kappa_3)$ and $\phi$ as marginals. The circular-linear density with copula (\ref{copcirlin:copula_new_family}) and $\alpha=(2\pi)^{-1}$ is given by
\[
p_3(\theta,x)=\left[1+\frac{1}{2\pi}\cos(2\pi\Psi_{\mathrm{vM}}(\theta;\mu_3,\kappa_3))(1-2\Phi(x))\right]\times\varphi_{\mathrm{vM}}(\theta;\mu_3,\kappa_3)\phi(x).
\]
\end{example}

\begin{example}[Reflected Frank copula with von Mises and normal marginal distributions]\label{copcirlin:p_example4}
Take $\varphi_{\mathrm{vM}}(\cdot;\mu_4,\kappa_4)$ and $\phi$ as marginals. The circular-linear density with reflected Frank copula ($\alpha=10$) is obtained by the mixture construction (\ref{copcirlin:copula_new_family2}):
\[
p_4(\theta,x)=c_{\Theta,X}^{\alpha}\lp \Psi_{\mathrm{vM}}(\theta;\mu_4,\kappa_4),\Phi(x) \rp\times\varphi_{\mathrm{vM}}(\theta;\mu_4,\kappa_4)\phi(x).
\]
\end{example}

As commented in Section 2, the formulation of the joint circular-linear density in terms of copulas simplifies the simulation of random samples. The general idea is to split the joint distribution $P$ by Sklar's theorem in a copula $C_{\Theta,X}$ and marginals $\Psi$ and $F$. Therefore, if we simulate a sample from $(U,V)$ (uniform random variables with copula $C_{\Theta,X}$) and we apply the marginal quantiles transformations, then $(\Psi^{-1}(U),F^{-1}(V))$ will be a sample from the distribution $P$. \\

The simulation of $(U,V)$ values from the copula $C_{\Theta,X}$ can be performed by the conditional method for simulating multivariate distributions (see \cite{Johnson1987}). The conditional distribution of $V$ given $U=u$, denoted by $C_u$, can be expressed as
\begin{equation}
C_u(v)=\frac{\partial C_{\Theta,X}(u,v)}{\partial u}=\int_0^v c_{\Theta,X}(u,t)\,dt,
\label{copcirlin:marginal_copula}
\end{equation}
where the first equality is an immediately property of copulas. So, for simulating random samples for the examples, or more generally, for simulating random samples of circular-linear random\nopagebreak[4] variables with density (\ref{copcirlin:circular_linear_copula}), we may proceed with the following algorithm.

\begin{figure}[h!]
	\centering
	\includegraphics[width=14cm]{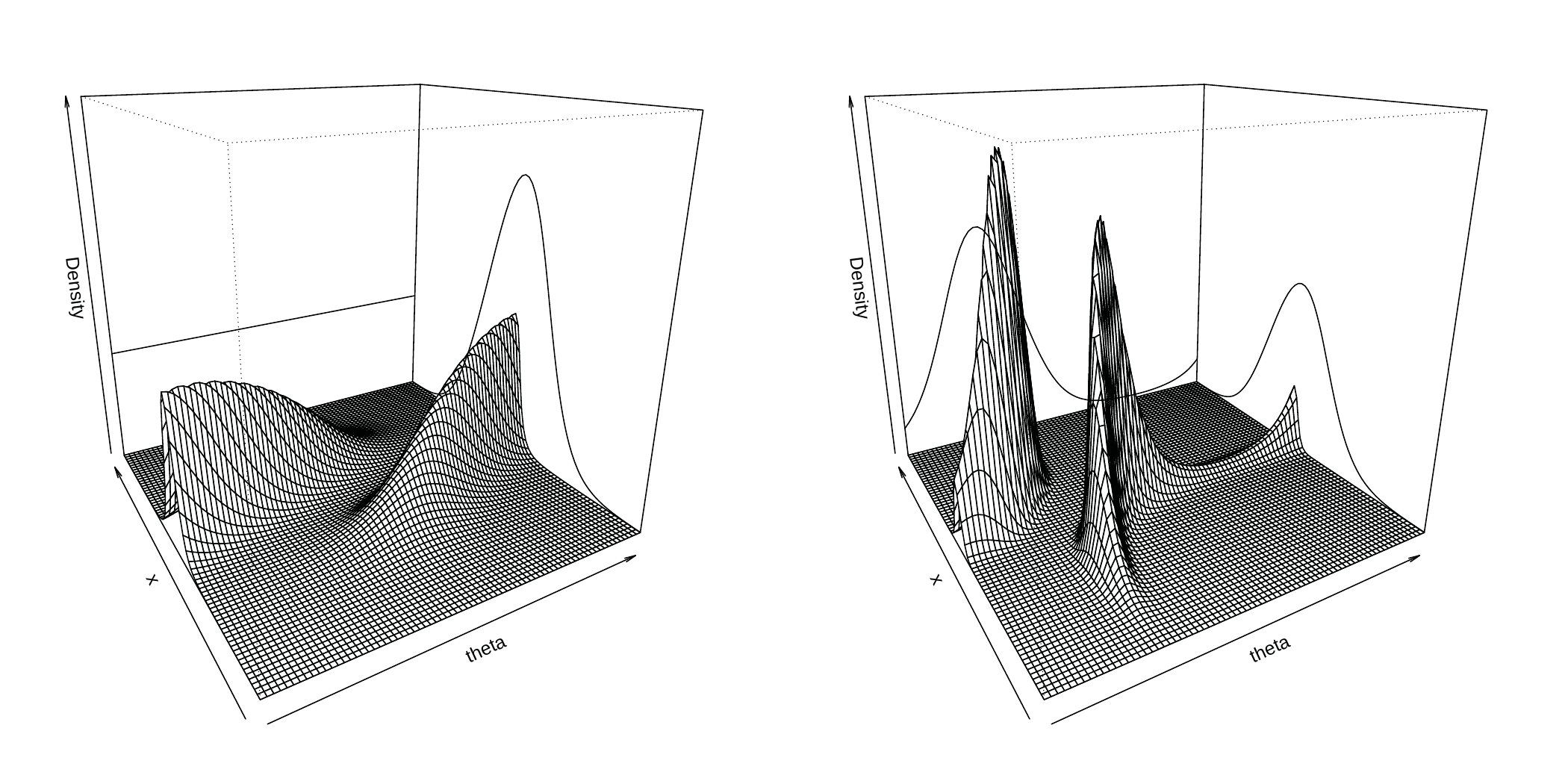}\\[-0.25cm]
	\includegraphics[width=14cm]{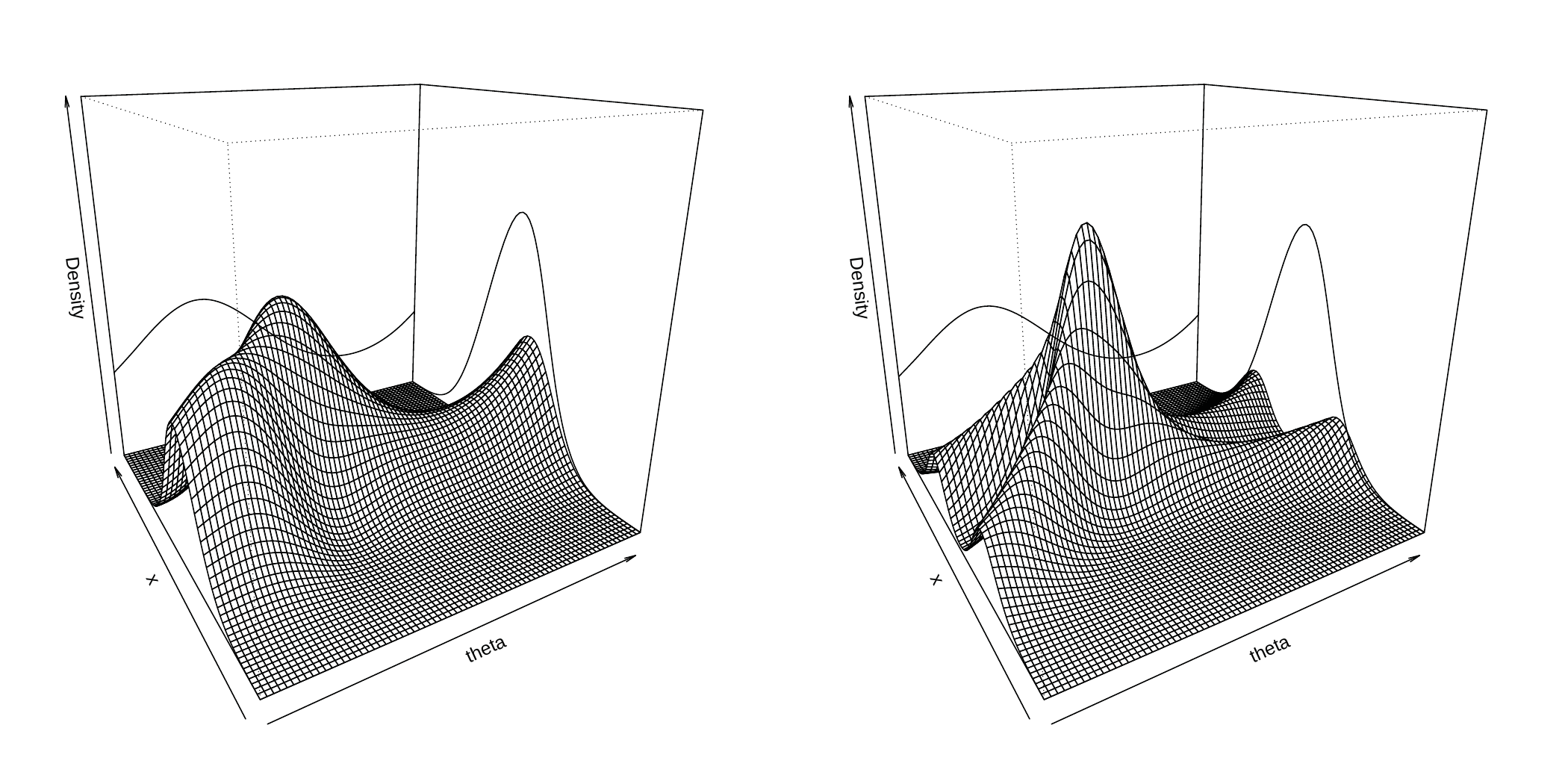}
	\caption{\small Density surfaces for the simulation study. Top left: Example \ref{copcirlin:p_example1} with $\mu=\pi$ and $\kappa=2$. Top right: Example \ref{copcirlin:p_example2} with $\mu'=\pi$, $\kappa'=5$, $\mu_2=\pi/2$ and $\kappa_2=2$. Bottom left: Example \ref{copcirlin:p_example3} with $\alpha=(2\pi)^{-1}$,  $\mu_3=\pi/2$ and $\kappa_3=0.5$. Bottom right: Example \ref{copcirlin:p_example4} with $\alpha=10$, $\mu_4=\pi/2$ and $\kappa_4=0.5$. \label{copcirlin:fig1}}
\end{figure}

\begin{algo}[Simulation algorithm]
\label{copcirlin:algo:2}\mbox{}
\begin{enumerate}[label=\textit{\roman{*}}., ref=\textit{\roman{*}}]
	\item Simulate $U,W\sim \mathcal{U}(0,1)$, where $\mathcal{U}(0,1)$ stands for the standard uniform distribution.\label{copcirlin:algo:2:1}
	\item Compute $V=C_u^{-1}\lp W \rp$.\label{copcirlin:algo:2:2}
	\item Obtain $\Theta=\Psi^{-1}(U)$, $X=F^{-1}(V)$.\label{copcirlin:algo:2:3}
\end{enumerate}
\end{algo}

In step \ref{copcirlin:algo:2:1}, two independent and uniformly distributed random variables are simulated. The conditional simulation method for obtaining $(U,V)$ from the circular-linear copula $C_{\Theta,X}$ is performed in step \ref{copcirlin:algo:2:2}. Finally, quantile transformations from the marginals are applied, obtaining a sample from $(\Theta,X)$ following the joint density (\ref{copcirlin:circular_linear_density}). For Examples \ref{copcirlin:p_example1} and \ref{copcirlin:p_example2}, the conditional distribution $C_u$ in (\ref{copcirlin:marginal_copula}) is related to a von Mises $\mathrm{vM}(\mu,\kappa)$ and in step \ref{copcirlin:algo:2:2}, for each $u$, $V=(2\pi)^{-1}\;\Psi_{\mathrm{vM}}^{-1}\lp W;2\pi u-\mu,\kappa\rp$. For Example \ref{copcirlin:p_example3}, for each $u$, $V=\frac{a+1-\sqrt{\lp a+1\rp^2-4a W}}{2a}$, with $a=2\pi\alpha\cos\lp 2\pi u\rp$. Example \ref{copcirlin:p_example4} is simulated from the mixture of copulas (\ref{copcirlin:copula_new_family2}), using that for the Frank copula, step \ref{copcirlin:algo:2:2} becomes $V=-\frac{1}{\alpha}\log\lp 1+\frac{W(1-e^{-\alpha})}{W(e^{-\alpha u}-1)-e^{-\alpha u}}\rp$.\\

The estimation algorithm proposed will be applied to estimate the densities in the examples. Different versions of the estimation algorithm can be implemented, considering parametric and nonparametric estimation methods. In addition, for the J\&W densities (Examples \ref{copcirlin:p_example1} and \ref{copcirlin:p_example2}), the estimation of the circular-linear density can be approached by representation (\ref{copcirlin:circular_linear_density}), in terms of a circular joining density, or by the more general representation (\ref{copcirlin:circular_linear_copula}). Summarizing, the following variants of the estimation algorithm will be presented:
\begin{enumerate}[label=\textit{\roman{*}}.]
\item J\&W, parametric (JWP): based on representation (\ref{copcirlin:circular_linear_density}), marginals as well as joining density are parametrically estimated. The results will be used as a benchmark for the J\&W models (Examples \ref{copcirlin:p_example1} and \ref{copcirlin:p_example2}).
\item J\&W, semiparametric (JWSP): based on representation (\ref{copcirlin:circular_linear_density}), marginals are estimated parametrically and a nonparametric kernel method is used for the joining density.
\item J\&W, nonparametric (JWNP): based on representation (\ref{copcirlin:circular_linear_density}), marginals and joining density are estimated by kernel methods.
\item Copula, semiparametric (CSP): based on representation (\ref{copcirlin:circular_linear_copula}), parametric estimation is considered for marginals. The copula density is estimated by kernel methods.
\item Copula, nonparametric (CNP): based on representation (\ref{copcirlin:circular_linear_copula}), marginals and copula density are estimated by kernel methods.
\end{enumerate}

In the parametric case, density estimators have been obtained by MLE, specifying the von Mises family for the circular distributions and the normal family for the linear marginal. Nonparametric estimation has been carried out using kernel methods. The kernel density estimator in (\ref{copcirlin:kernel_linear}), with Gaussian kernel and \cite{Sheather1991} bandwidth, has been used for obtaining $\hat f$. For $\hat\varphi$ and $\hat g$, the circular kernel density (\ref{copcirlin:kernel_circular}) has been implemented, with exponential kernel and likelihood cross-validatory bandwidth. In the semiparametric approaches (parametric marginals and nonparametric joining density or copula), Maximum Likelihood has been used for obtaining $\hat\varphi$ and $\hat f$. The circular kernel estimator (\ref{copcirlin:kernel_circular}) has been considered for $\hat g$. The copula density kernel estimator (\ref{copcirlin:np_copula}) has been used for $\hat c_{\Theta,X}$, with bivariate Gaussian kernel and restricted bandwidth matrix. Specifically, following the procedure of \cite{Duong2003}, full bandwidth matrices have been tried, but with non significant differences in the values of the principal diagonal. Hence, a restricted bandwidth with two smoothing parameter values is used, considering the same element in the principal diagonal and a second element in the secondary diagonal, regarding for the kernel orientation.\\

In order to check the performance of the procedure for estimating circular-linear densities, the MISE criterion is considered:
\[
\mbox{MISE}=\int_0^{2\pi}\int_{-\infty}^\infty \mathbb{E}\lrc{(\hat p(\theta,x)-p(\theta,x))^2}\,dx\,d\theta.
\]
The MISE is approximated by Monte Carlo simulations, taking $1000$ replicates. Four sample sizes have been used: $n=50$, $n=100$, $n=500$ and $n=1000$. In the first example, the set-up parameters are $\mu=\pi$ and $\kappa=2$. For the second example, we take $\mu'=\pi$, $\kappa'=5$, $\mu_2=\pi/2$ and $\kappa_2=2$. Both in Examples \ref{copcirlin:p_example3} and \ref{copcirlin:p_example4} the parameters in the von Mises marginal were set to $\mu_3=\mu_4=\pi/2$ and $\kappa_3=\kappa_4=0.5$, and the linear marginal is a standard normal.\\

In Figure \ref{copcirlin:fig1}, surface plots for the example densities are shown, the top row corresponding to the J\&W family (Example \ref{copcirlin:p_example1} and Example \ref{copcirlin:p_example2}). Simulation results for these examples can be seen in Table \ref{copcirlin:MISE_table1}. For all the alternatives of the algorithm, the MISE is reduced when increasing the sample size. Example \ref{copcirlin:p_example2} presents higher values for the MISE, and it is due to the estimation of a more complex structure in the circular marginal density (circular uniform in Example \ref{copcirlin:p_example1} and von Mises in Example \ref{copcirlin:p_example2}). In both models, the estimation methods providing the information about the J\&W structure (that is, based on representation (\ref{copcirlin:circular_linear_density})) work better, as expected. Nevertheless, the copula based approaches, CSP and CNP, are competitive with the JWNP.\\

\begin{figure}[h!]
\centering
		\includegraphics[trim=0 0 0 1.5cm,clip,width=14cm]{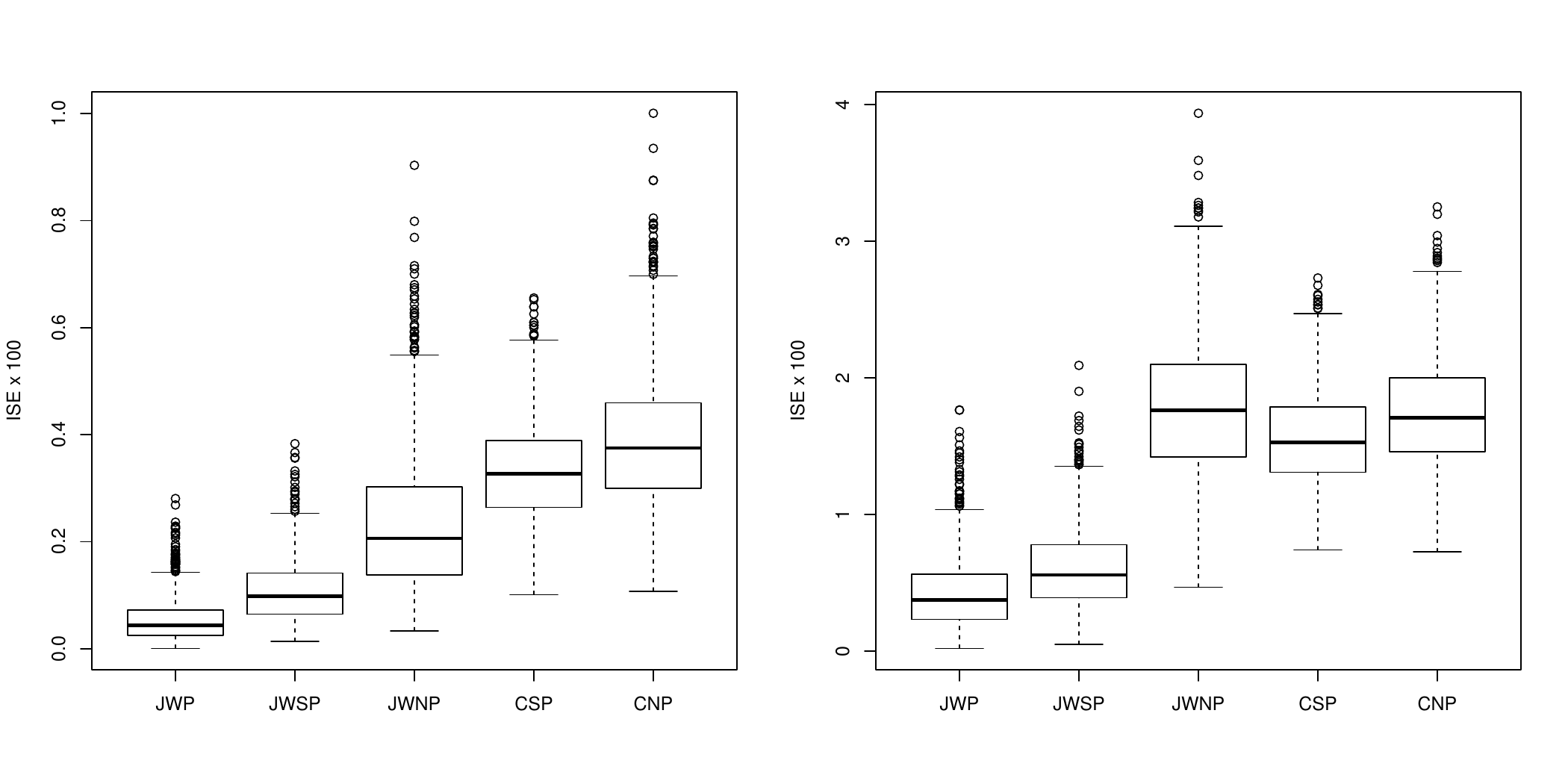}\\[0.25cm]
		\includegraphics[trim=0 0 0 1.5cm,clip,width=14cm]{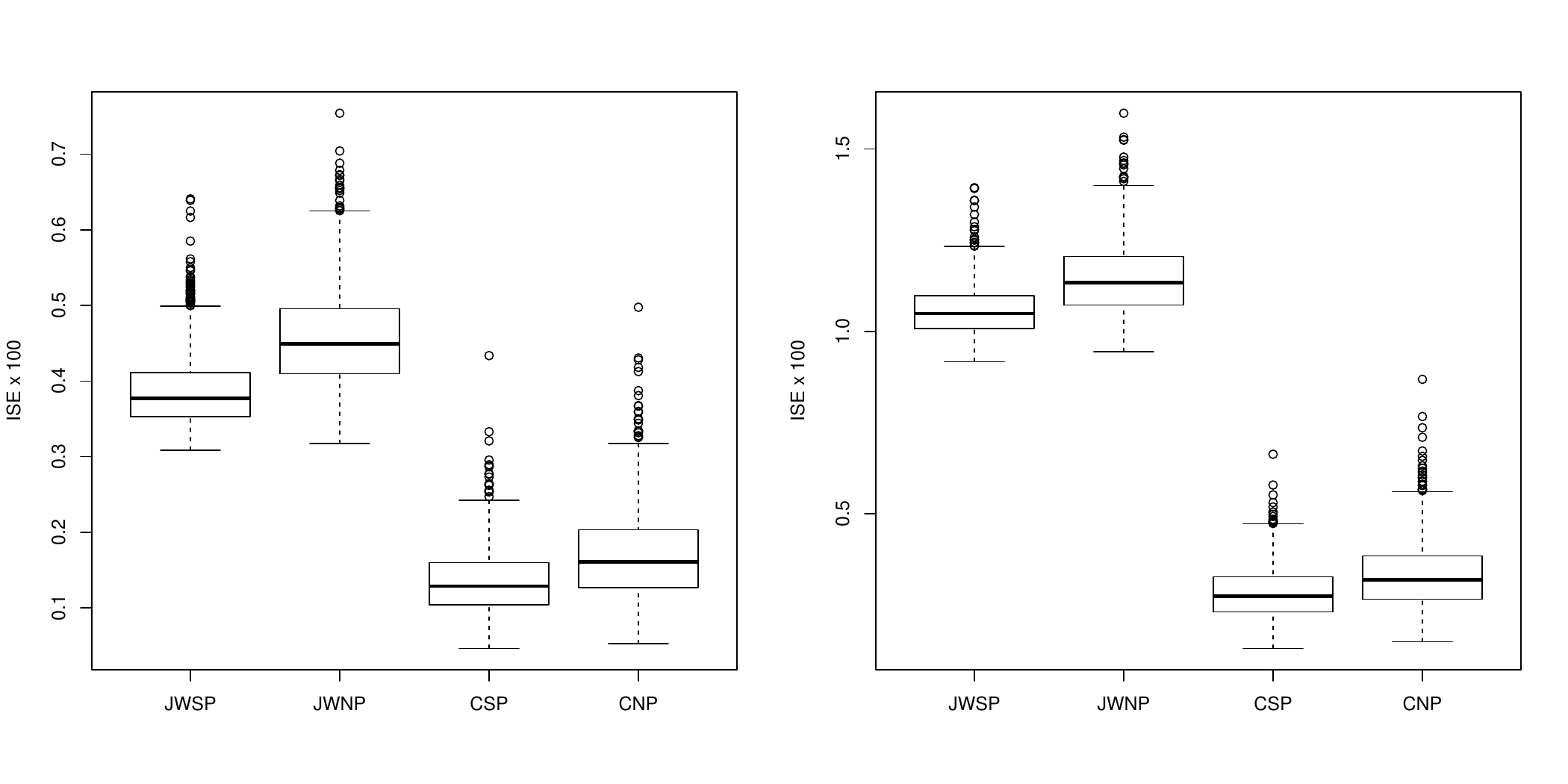}
		\caption{\small Boxplots of the $\mathrm{ISE}\times100$ for Example \ref{copcirlin:p_example1} (top left), Example \ref{copcirlin:p_example2} (top right), Example \ref{copcirlin:p_example3} (bottom left) and Example \ref{copcirlin:p_example4} (bottom right) for $n=500$ and different estimation procedures.}
		\label{copcirlin:boxplot}
\end{figure}

The parametric method JWP presents the lowest MISE values for all sample sizes in both examples, so it will be taken as a benchmark for computing the relative efficiencies of the nonparametric and semiparametric approaches, both based on the density (\ref{copcirlin:circular_linear_density}) or on the copula density (\ref{copcirlin:circular_linear_copula}). Relative efficiencies are obtained as the ratio between the MISE of the parametric method and the MISE of the nonparametric and semiparametric procedures. The relative efficiencies (see Table \ref{copcirlin:MISE_table1}) are higher for the semiparametric approach, with better results for Example \ref{copcirlin:p_example2}. Boxplots for the ISE for sample size $n=500$ can be seen in Figure \ref{copcirlin:boxplot}, top row. The larger variability of the nonparametric methods (JWNP and CNP) can be appreciated for both examples. For Example \ref{copcirlin:p_example2} (see Figure \ref{copcirlin:boxplot}, top right plot), note that JWNP shows the highest values for the ISE. 

\begin{table}[!h]
\centering
\small
\begin{tabular}{ll|ccc|cc||cccc}\toprule\toprule
\multicolumn{2}{c}{} &  \multicolumn{3}{c}{J\&W-based} & \multicolumn{2}{c}{Copula-based} & \multicolumn{4}{c}{Relative efficiency} \\\midrule
&$n$& JWP & JWSP & JWNP & CSP & CNP & JWSP & JWNP & CSP & CNP \\
\midrule
Example \ref{copcirlin:p_example1} & $50$ & $0.534$ & $0.741$ & $1.395$ & $1.417$ & $1.851$ & $0.721$ & $0.383$ & $0.377$ & $0.289$\\ 
& $100$ & $0.266$ & $0.420$ & $0.851$ & $0.923$ & $1.180$ & $0.632$ & $0.312$ & $0.288$ & $0.225$\\ 
& $200$ & $0.132$ & $0.234$ & $0.485$ & $0.596$ & $0.735$ & $0.563$ & $0.272$ & $0.221$ & $0.179$\\ 
& $500$ & $0.055$ & $0.109$ & $0.237$ & $0.333$ & $0.392$ & $0.504$ & $0.232$ & $0.165$ & $0.140$\\ 
& $1000$ & $0.027$ & $0.062$ & $0.136$ & $0.216$ & $0.244$ & $0.432$ & $0.196$ & $0.124$ & $0.109$\\\midrule 
Example \ref{copcirlin:p_example2} & $50$ & $4.059$ & $4.671$ & $8.311$ & $7.497$ & $8.239$ & $0.869$ & $0.488$ & $0.541$ & $0.493$\\ 
& $100$ & $2.090$ & $2.516$ & $5.602$ & $4.845$ & $5.376$ & $0.830$ & $0.373$ & $0.431$ & $0.389$\\ 
& $200$ & $1.068$ & $1.362$ & $3.442$ & $3.015$ & $3.362$ & $0.784$ & $0.310$ & $0.354$ & $0.318$\\ 
& $500$ & $0.429$ & $0.610$ & $1.797$ & $1.566$ & $1.746$ & $0.703$ & $0.239$ & $0.274$ & $0.246$\\ 
& $1000$ & $0.211$ & $0.336$ & $1.061$ & $0.941$ & $1.046$ & $0.627$ & $0.199$ & $0.224$ & $0.202$\\\bottomrule\bottomrule 
\end{tabular}
\caption{\small $\mathrm{MISE}\times100$ for estimating the circular-linear density in Examples \ref{copcirlin:p_example1} and \ref{copcirlin:p_example2}. Relative efficiencies for JWSP, JWNP, CSP and CNP are taken with respect to JWP.}
\label{copcirlin:MISE_table1}
\end{table}
\begin{table}[!h]
\centering
		\small
		\begin{tabular}{ll|cc|cc}\toprule\toprule
			\multicolumn{2}{c}{} &  \multicolumn{2}{c}{J\&W-based} & \multicolumn{2}{c}{Copula-based} \\
			\midrule
			&$n$&  JWSP & JWNP & CSP & CNP  \\
			\midrule
Example \ref{copcirlin:p_example3} & $50$ & $0.881$ & $1.150$ & $0.612$ & $0.826$\\ 
& $100$ & $0.625$ & $0.813$ & $0.372$ & $0.506$\\ 
& $200$ & $0.483$ & $0.595$ & $0.237$ & $0.307$\\ 
& $500$ & $0.389$ & $0.459$ & $0.135$ & $0.171$\\ 
& $1000$ & $0.357$ & $0.403$ & $0.091$ & $0.109$\\\midrule 
Example \ref{copcirlin:p_example4} & $50$ & $1.648$ & $1.995$ & $1.158$ & $1.443$\\ 
& $100$ & $1.339$ & $1.568$ & $0.747$ & $0.926$\\ 
& $200$ & $1.168$ & $1.322$ & $0.502$ & $0.607$\\ 
& $500$ & $1.058$ & $1.148$ & $0.284$ & $0.333$\\ 
& $1000$ & $1.019$ & $1.075$ & $0.183$ & $0.210$\\\bottomrule\bottomrule 
		\end{tabular}
		\caption{\small $\mathrm{MISE}\times100$ for estimating the circular-linear density in Examples \ref{copcirlin:p_example3} and \ref{copcirlin:p_example4}.}
		\label{copcirlin:MISE_table2}
\end{table}

Obviously, the first three proposals (JWP, JWSP, JWNP) make sense for distributions belonging to the J\&W family. However, since the marginals considered in Examples \ref{copcirlin:p_example3} and \ref{copcirlin:p_example4} also belong to the von Mises (circular) and the normal (linear) classes, as in Examples \ref{copcirlin:p_example1} and \ref{copcirlin:p_example2}, JWSP, JWNP, CSP and CNP variants of the algorithm will be applied in the last two cases. Results are reported in Table \ref{copcirlin:MISE_table2}. MISE values decrease with sample size, showing CSP and CNP a similar behaviour. Note also that assuming a representation like (\ref{copcirlin:circular_linear_density}), increases dramatically the MISE: for instance, in Example \ref{copcirlin:p_example3}, for $n=500$ or $n=1000$, the MISE for JWSP or JWNP is four times the one provided by CNP. The effect of this misspecification in the copula structure can be clearly seen in the bottom row of Figure \ref{copcirlin:boxplot}. 

\section{\texorpdfstring{Application to wind direction and SO$_2$ concentration}{Application to wind direction and SO2 concentration}}
\label{copcirlin:application}

The goal of this work is to explore the relation between wind incidence direction and  SO$_2$ concentration in monitoring station B2 near a power plant (see Figure \ref{copcirlin:locations} for location of station B2). SO$_2$ is measured in $\mu g/m^3$ and wind direction as a counterclockwise angle in $[0,2\pi)$. With this codification, $0$, $\frac{\pi}{2}$, $\pi$ and $\frac{3\pi}{2}$ represent east, north, west and south direction, respectively.\\

The dataset contains observations recorded minutely in January 2004 and January 2011, but due to technical limitations in the measuring device, SO$_2$ is only registered when it is higher than $3\,\mu g/m^3$. Concentration values below this threshold are considered as non significant and are recorded as the lower detection limit ($3\,\mu g/m^3$). Data have been hourly averaged, resulting $736$ observations for 2004 and $743$ for 2011. In order to avoid repeated data, perturbation procedures have been applied to both marginals, and will be detailed below. Afterwards, a Box--Cox transformation for the SO$_2$ concentration with $\lambda=-0.84$ for 2004 and $\lambda=-7.34$ for 2011 have been applied. For the sake of simplicity, we will refer to these transformed data as SO$_2$ concentration, but note that figures are shown in the transformed scale.\\

Measurement devices, both for the wind direction and for SO$_2$ concentration, did not present a sufficient precision to avoid repeated data, and this problem was inherited also for the hourly averages. The appearance of repeated measurements posed a problem in the application of the procedure, specifically, in the bandwidth computation. Perturbation in the linear variable, the SO$_2$ concentration, was carried out following \cite{Azzalini1981}. A pseudo-sample of SO$_2$ levels is obtained as\nolinebreak[4] follows:
\[
\widetilde X_i=X_i+b\varepsilon_i,
\]
where $X_i$ denote the observed values, $b=1.3\hat\sigma n^{-1/3}$ and $\varepsilon_i$, $i=1,\ldots,n$, are independent and identically distributed random variables from the Epanechnikov kernel in $(-\sqrt 5,\sqrt 5)$. $\hat\sigma$ is a robust estimator of the variance, which has been computed using the standardized interquartile range. \cite{Azzalini1981} showed that this choice of $b$ for the data perturbation allows for consistent estimation of the distribution function, getting a mean squared error with the same magnitude as the one from the empirical cumulative distribution function.\\

The problem of repeated measures also occurs for wind direction. In this case, a perturbation procedure similar to the linear variable case can be used, just considering the circular variable (the angle) in the real line. Then, a pseudo-sample of wind direction was obtained as
\[
\widetilde\theta_i=\theta_i+d\varepsilon_i,
\]
with $\theta_i$ denoting the wind direction measurements and $\varepsilon_i$, $i=1,\ldots,n$, were independently generated from a von Mises distribution with $\mu=0$ and $\kappa=1$, with $d=n^{-1/3}$. We have checked by simulations that the applied perturbation did not affect the distribution of the data.

\subsection{\texorpdfstring{Exploring SO$_2$ and wind direction in 2004 and 2011}{Exploring SO2 and wind direction in 2004 and 2011}}
\label{copcirlin:exploring}

\begin{figure}[h!]
	\vspace*{-0.25cm}
	\centering
	\includegraphics[width=14cm]{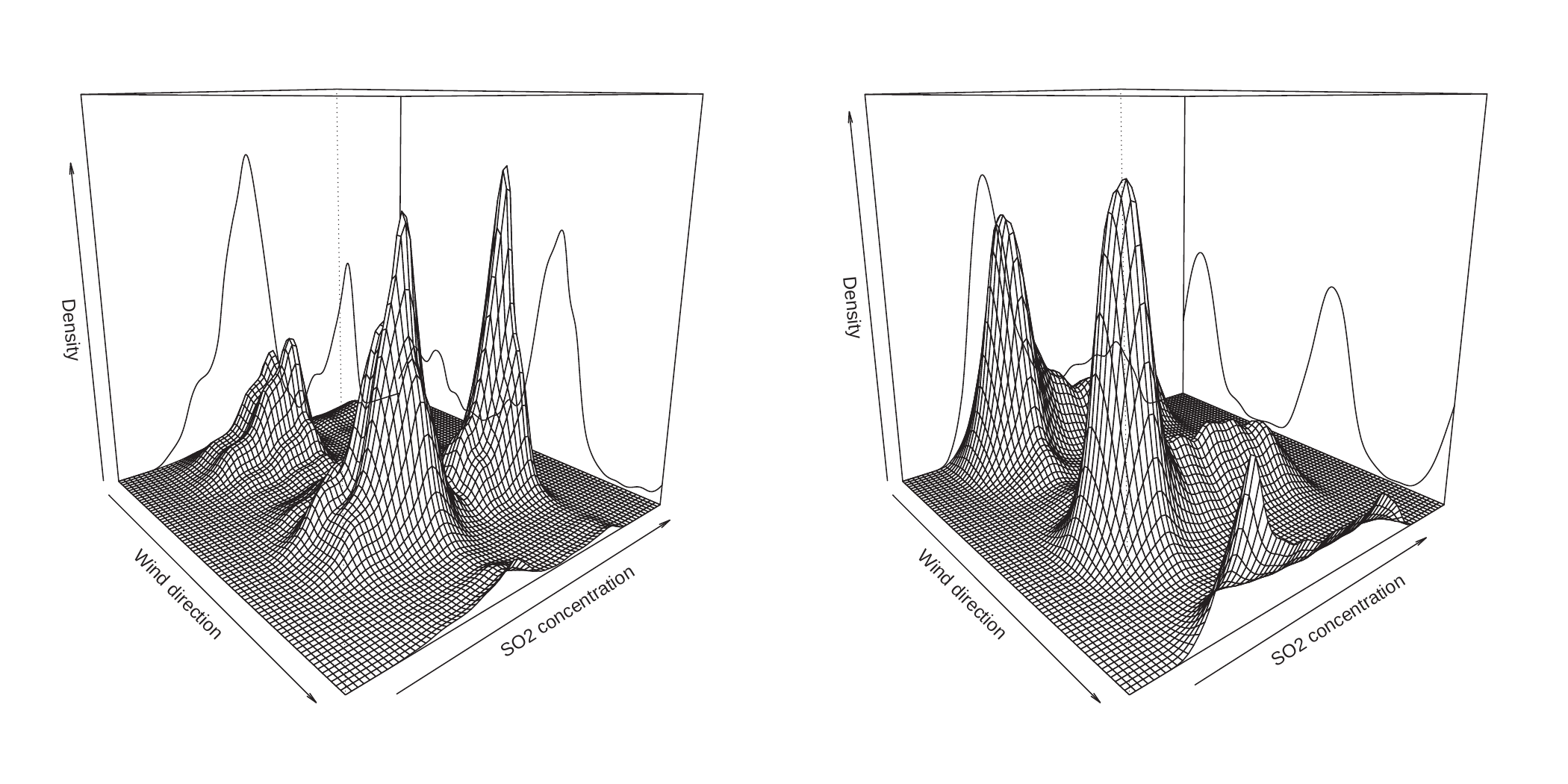}\\[-0.25cm]
	\includegraphics[width=14cm]{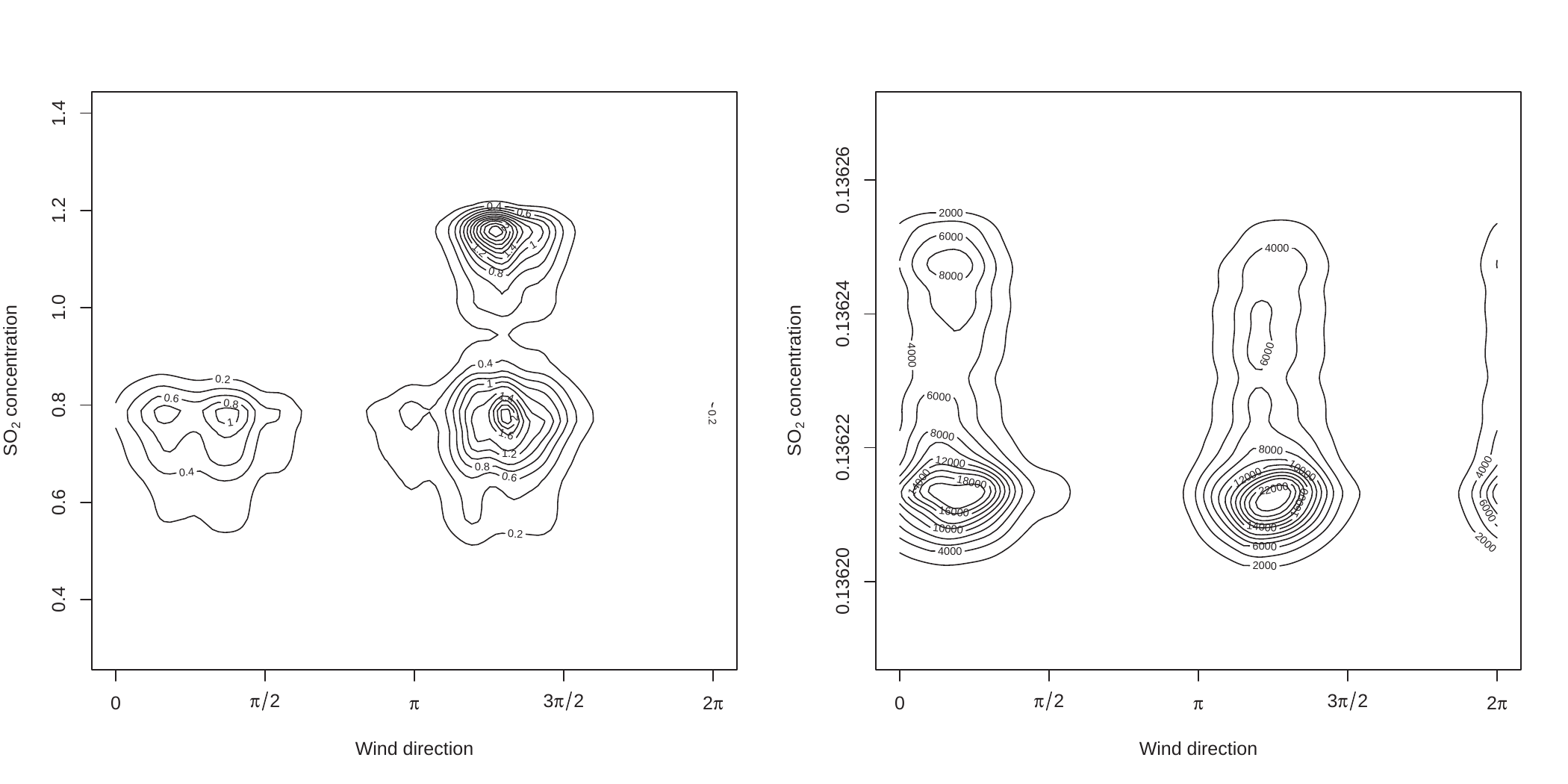}
	\caption{\small Circular-linear density estimator for wind direction and SO$_2$ concentrations in monitoring station B2. Left column: 2004. Right column: 2011.}
	\label{copcirlin:B2_surface}
\end{figure}

The estimation procedure is applied to data from 2004 and 2011 in station B2, considering the fully nonparametric approach. Specifically, a nonparametric kernel density estimator for the SO$_2$ concentration was used, with the plug-in rule bandwidth obtained by method of \cite{Sheather1991} ($h=1.75\times 10^{-6}$ for 2011, $h=0.02$ for 2004). For the wind direction, circular kernel density estimation has been also performed, with likelihood cross-validatory bandwidth ($\nu=46.41$ for 2011, $\nu=78.56$ for 2004). In Figure \ref{copcirlin:B2_surface}, the estimation of the circular-linear density surfaces, with the corresponding contour plots, is shown. \\

For 2011 two modes in the SW and NE directions (see Figure \ref{copcirlin:B2_surface}, right column) can be identified, both with a similar behaviour and with quite low values of SO$_2$ concentrations. Recall that the scale is Box--Cox transformed (for data in the original scale, see Figure \ref{copcirlin:diagrams}). A different situation occurs in 2004 (see Figure \ref{copcirlin:B2_surface}, left column) where, in addition to the two modes that appear in 2011 for low values of SO$_2$, a third mode arises. This mode is related to winds blowing from SW (from the thermal power plant) and to SO$_2$ values significantly higher than for 2011. This relation suggests that the additional mode represents SO$_2$ pollutants coming from the power plant, and its disappearance for 2011 illustrates the effectiveness of the control measures applied during 2005--2008 to reduce the SO$_2$ emissions from the power plant.

\section{Final comments}
\label{copcirlin:final}

A flexible algorithm for estimating circular-linear densities is proposed based on a copula density representation. The method provides a completely nonparametric estimator, but it can be modified to accommodate the classical Johnson and Wehrly family of circular-linear distributions. In the purely nonparametric version of the algorithm, circular and linear kernel density estimators have been used for the marginals, although other nonparametric density estimators could be considered. In addition, the extension of the algorithm for circular-circular density estimation is straightforward.\\

In our air quality data application the precision of the measurement devices posed some extra problems in the data analysis. The lack of precision resulted in the appearance of repeated values for the wind direction, and a data perturbation procedure was needed in order to apply the algorithm. The perturbation method proposed has been checked empirically, and it is inspired by the results for kernel distribution estimation, but our guess is that similar results could be obtained with just perturbing the data by summing errors from a highly concentrated distribution (e.g. a von Mises distribution with large $\kappa$). Nevertheless, data perturbation in the circular setting needs further investigation. Another possible problem that may be encountered in practice, for linear variables, is censoring, that may be due to detection limits or other phenomena. Under censoring, the observation values are only partially known, and suitable estimation procedures for density estimation with censored data should be applied.

Finally, a natural criticism of our analysis of the air quality data is that the measurements were not independent of one another. Temporal dependence could be accounted for in the estimation procedure by using a proper bandwidth selector, such as a $k$-fold cross-validatory bandwidth for the linear kernel density estimator. However, there are not such alternatives for circular data (up to the authors' knowledge) and the study of bandwidth selection rules for circular dependent data is beyond the scope of this paper.\\

The simulation study and real data analysis has been carried out in \texttt{R} 2.14 (\cite{RDevelopmentCoreTeam2011}), using self-programmed code and package \texttt{circular} (\cite{circularold}). For the real data analysis, the computing time for 2004 is $30.77$ seconds, taking the computation of the copula estimator $8.58$ seconds. The same procedures take $31.57$ seconds for the 2011 data. All computations were done on a computer with $1.6$ GHz core. This shows that the computational cost of the method is not high and its application is feasible in practice.

\section*{Acknowledgements}

The authors acknowledge the support of Project MTM2008-03010, from the Spanish Ministry of Science and Innovation and Project 10MDS207015PR from Direcci\'on Xeral de I+D, Xunta de Galicia. Work of E. Garc\'ia-Portugu\'es has been supported by FPU grant AP2010-0957 from the Spanish Ministry of Education. We also thank the referee and the Associate Editor for providing constructive comments and help in improving this paper.

\end{document}